\documentclass[prb,aip,aps,graphicx,twocolumn]{revtex4}
\usepackage{graphicx}
\usepackage{amssymb}
\usepackage{subfigure}
\usepackage{color}
\begin{document}
\title{Structural ordering driven anisotropic magnetoresistance, anomalous Hall resistance and its
topological overtones in full-Heusler Co$_2$MnSi thin films}

\author{Himanshu Pandey$^1$ and R. C. Budhani$^{1,2^{\ast}}$}

\affiliation{$^1$Condensed Matter-Low Dimensional Systems Laboratory, Department of Physics,
Indian Institute of Technology, Kanpur-208016, India.
\\$^2$National Physical Laboratory, Council of Scientific and Industrial Research, New Delhi - 110012,
India.}

\email{rcb@iitk.ac.in, rcb@nplindia.org}
\date{\today}

\begin{abstract}

We report the evolution of crystallographic structure, magnetic
ordering and electronic transport in thin films of full-Heusler
alloy Co$_2$MnSi deposited on (001) MgO with annealing
temperatures ($T_A$). By increasing the $T_A$ from 300$^\circ$C to
600$^\circ$C, the film goes from a disordered nanocrystalline
phase to $B2$ ordered and finally to the $L2_1$ ordered alloy. The
saturation magnetic moment improves with structural ordering and
approaches the Slater-Pauling value of $\approx 5.0 \mu_B$ per
formula unit for $T_A$ = 600$^\circ$C. At this stage the films are
soft magnets with coercive and saturation fields as low as
$\approx$ 7 mT and 350 mT, respectively. Remarkable effects of
improved structural order are also seen in longitudinal
resistivity ($\rho_{xx}$) and residual resistivity ratio. A model
based upon electronic transparency of grain boundaries illucidates
the transition from a state of negative $d\rho/dT$ to positive
$d\rho/dT$ with improved structural order. The Hall resistivity
($\rho_{xy}$) derives contribution from the normal scattering of
charge carriers in external magnetic field, the anomalous effect
originating from built-in magnetization and a small but distinct
topological Hall effect in the disordered phase. The carrier
concentration ($n$) and mobility ($\mu$) have been extracted from
the high field $\rho_{xy}$ data. The highly ordered films are
characterized by $n$ and $\mu$ of 1.19$\times$ 10$^{29}$ m$^{-3}$
and 0.4 cm$^2V^{-1}s^{-1}$ at room temperature. The dependence of
$\rho_{xy}$ on $\rho_{xx}$ indicates the dominance of skew
scattering in our films, which shows a monotonic drop on raising
the $T_A$. The topological Hall effect is analyzed for the films
annealed at 300$^\circ$C. We find maximum topological contribution
to Hall resistivity around 0.5 T while it approach to zero with
increasing magnetic field. The anisotropic magnetoresistance
changes its sign from positive to negative as we go from as
deposited to 600$^\circ$C annealed film suggesting a gradual
increase in the half-metallic character.
\end{abstract}

\maketitle
\section{INTRODUCTION}
Although discovered a century ago by F. Heusler, the compounds
which bear his name have captured much attention in recent years
due to their fascinating properties which include
magneto-stuctural,\cite{Kainuma} magneto-optical,\cite{Krenke} and
magneto-caloric phenomena,\cite{Engen} large
thermo-electricity\cite{Bosu} and heavy fermion
superconductivity.\cite{Wernick} Recently, a new property has been
predicted in which spin-polarized edge and surface states are
topologically protected against impurity scattering, known as
topological insulator.\cite{Chadov,Lin} The coexistence of
superconductivity and a magnetically ordered state have also been
reported for Heusler alloys containing rare earth
elements.\cite{Kierstead} A certain class of full-Heusler alloys
are particularly important for spintronics applications. The
spin-polarized tunneling, which incorporates the imbalance between
the density of states (DOS) of two spin states in the conduction
band, controls the performance of devices like magnetic tunnel
junctions and spin valves, and can be improved significantly if
the conduction electrons are fully spin-polarized. These materials
are characterized by a spin-split conduction band with a clear
energy gap for the up and down spin states, and the Fermi level
($\varepsilon_F$) being located in one of them. This property of
the electronic structure is called half-metallicity and a large
number of Heusler alloys have been predicted to be half-metals.
Although there are other compounds such as
La$_{1-x}$Sr$_{x}$MnO$_{3}$ \cite{Park} and CrO$_2$\cite{Yji} with
measured spin polarization $> 90\%$ at low temperatures, but they
are not suitable for ambient temperature applications because of
their low ($<400$ K) Curie temperature ($T_C$). The Heusler
alloys, which have very high $T_C$, are much suitable for
spintronics applications, provided the spin-polarized nature of
the conduction band can be retained in the films of these
materials. However, due to atomic and/or antisite disorder, the
degree for spin polarization in epitaxial films is found to be
smaller than those obtained from various band structure
calculations. Additionally, there is a reduction in spin
polarization at the surfaces due to surface
reconstruction,\cite{Ristoiu} and disorder\cite{Kautzky} present
between different sublattices. Therefore, studies of atomic and
antisite disorder, surface reconstruction and spin polarization in
thin films of Heusler alloys is a challenging but worthwhile
problem to pursue.

Amongst all Heusler alloys, the Co-based full-Heusler compounds
are attractive due to their high $T_C$ and high magnetic moment
per formula unit (f.u.). The full-Heusler alloy Co$_2$MnSi (CMS)
belongs to the same family with $T_C$ = 985 K\cite{Fujii} and a
large minority-spin band gap of 0.4 eV. While, a large number of
full-Heusler alloy thin films have been grown on semiconductors
like Si,\cite{Zander} Ge,\cite{Kasahara} and GaAs;\cite{Wang}
there are few reports on the films grown on oxide substrates such
as Al$_2$O$_3$,\cite{Singh,Schneider}
MgO\cite{Schneider,Pandey,Schneider2} and
SrTiO$_3.$\cite{Anupam,Himanshu1} Various deposition techniques
such as laser
ablation,\cite{Wang,Schneider,Pandey,Anupam,Himanshu1,Rout}
sputtering\cite{Geiersbacha,Raphael,Kammerer,Kim} and molecular
beam epitaxy\cite{Ambrose} are in use for the growth of such films
under varying conditions of growth. It is well-known that the
choice of the deposition temperature, growth conditions, and
substrates can significantly alter the structural, magnetic, and
electronic properties of thin films.

Here our focus is to carryout an in-depth study of the electronic
transport and magnetism of CMS films grown on (001) MgO by a
single target pulsed laser ablation technique (PLD). The lattice
parameter ($ a_{bulk}\approx$ 0.5655 nm) of CMS matches quite well
with the face diagonal ($\sqrt 2 a_{sub}$) of MgO. The lattice
misfit [$(a - \sqrt 2 a_{sub} )/\sqrt 2 a_{sub}$] between the CMS
and MgO is $\approx -5\%$. It should be noted that the first
principle calculations by Kandpal \textit{et al.}\cite{Kandpal}
show that the strain $\approx$ $\pm$5\% does not alter the band
structure significantly. Thus, the charge transport in these films
should behave in the same way as that of the bulk samples. The
fabrication of stoichiometric films from an alloy target is the
key strength of PLD. This technique relies on the formation of a
local superheated region on the target, which then explodes as a
plasmonic plume maintaining the stoichiometry of the target. The
high thermal conductivity of elemental metals, however, hinders
the formation of a superheated spot on the target and the material
is removed by splashing, rather than ablation. Therefore, the
parameters such as laser energy density, pulse repetition rate,
and target-to-substrate distance need to be optimized very
carefully.

The half-metallic character and saturation magnetization of
Heusler alloys are linked to atomic structure and crystallographic
ordering in the material. In thin films, the optimum realization
of these two parameters is affected by lattice strain, interfacial
metallurgy and surface roughness. The Heusler compound CMS
crystallizes in the cubic space group $Fm\bar{3}m$ with
$\textit{L}2_1$ type ordering. This structure can be built from a
zinc blende-type sublattice of one Co and one Si. The second Co
fills the tetrahedral holes of the structure whereas Mn occupies
the octrahedral holes. Theoretical calculations show that a small
variation from this ordered arrangement of atoms can significantly
affect the electronic and magnetic properties of the Heusler
compounds. The greater atomic ordering is realized when the
compound is synthesized at high temperatures. For electronic
applications, it is desirable to grow films at as low a
temperature as possible to ensure high surface smoothness and
integration with other device process steps. Here we report the
evolution of atomic ordering in CMS films deposited on (001) MgO
at 200$^\circ$C and subsequently annealed at several temperatures.
The extent of atomic ordering has been monitored by two
magneto-transport measurements; namely Anisotropic
magnetoresistance (AMR) and Hall effect, which respectively yield
quantitative information about the degree of spin polarization,
saturation magnetization (M$_S$) and spin chirality. These data
have been supported by direct measurement of magnetization,
longitudinal resistivity and crystallographic structure. This
paper comprises of following sections, out of which the
introduction has been presented already in Sec. I. The sample
preparation and various measurement procedures are given in Sec.
II. All the results and their discussions are given in Sec. III,
which deals with the structural and elemental characterization in
addition to magnetic ordering and electronic transport in the CMS
films. The anomalous Hall effect (AHE), topological Hall effect
(THE), and AMR results are described in Sec. IV and finally, the
conclusions drawn from these studies are given in Sec. V.

\section{EXPERIMENTAL DETAILS}
40 nm thick films of CMS were deposited by pulsed laser ablation
of stoichiometric target of the alloy. A KrF excimer laser
(wavelength = 248 nm and pulse width $\approx$ 20 ns) was used for
the film deposition. The target was prepared by arc melting a
mixture of the constituents (99.99\% pure) in argon atmosphere.
The alloy ingot was re-melted many times and then annealed at
1000$^0$C for 24 hours to secure homogenization. Prior to the
annealing, the target was wrapped with tantalum foil in order to
avoid any undesired oxidation and then placed in a quartz tube,
which was subsequently evacuated and sealed. The resulting target
has \emph{L}$2_1$ structure and elemental stoichiometry in the
ratio of 2:1:1 for Co, Mn, and Si, respectively as confirmed by
X-ray diffraction (XRD) and energy dispersive X-ray analysis
(EDAX). A Perfect site order can usually be achieved by preparing
the films on lattice matched substrates at high temperatures. This
is often undesirable for devices where requirement of smooth
surfaces, sharp interfaces and minimized inter-diffusion are the
key for performance. We used a growth temperature of 200$^\circ$C
followed by one hour annealing at $T_A$ = 300, 400, 500, and
600$^\circ$C to enhance crystallization and ordering in the films.
The growth and post-deposition annealing were carried out in an
all-metal-seal high vacuum chamber equipped with a Ti-getter pump
and an ultra high vacuum compatible substrate heater which could
heat the samples to $\approx$ 850$^\circ$C. Prior to the thin film
deposition, the MgO substrates were annealed at 800$^\circ$C for
30 minutes to allow surface reconstruction. Afterwards, the
temperature was lowered to 200$^\circ$C to commence the growth.
The crystallographic structure of the films was characterized in
$\theta$-2$\theta$, $\omega$, $\varphi$, and grazing incidence
X-ray diffraction (GIXRD) modes using a PANalytical X'Pert PRO
X-ray diffractometer equipped with a Cu$K_{\alpha1}$ source. From
the simulation of X-ray reflectivity (XRR) curves by a genetic
algorithm, we estimated the film thickness. The surface roughness
of the films were obtained by XRR and atomic force microscopy
(AFM). The high resolution field emission scanning electron
microscopy (FESEM, Model: SUPRA40VP) and EDAX were used to check
the homogeneity, stoichiometry and uniformity of the thin films.
For room temperature magnetization measurements with both in-plane
and out-of-plane field orientations, we have used a vibrating
sample magnetometer. For the electrical measurements, samples were
patterned in a four-probe geometry using $Ar^+$ ion milling
through a stainless steel shadow mask and the silver pads for
electrical contact were deposited by thermal evaporation. The
resistivity and AMR measurement were performed in constant current
mode using precision programmable dc current source (Keithley
224), digital temperature controller (Scientific Instruments Inc
model 9650), and nanovoltmeter (Keithley 2182). For Hall
measurement, samples were patterned into standard six probe Hall
bar geometry and measured in Physical Property Measurement System
(PPMS, Quantum Design).

\section{RESULTS AND DISCUSSION}
\subsection{Structural and elemental characterization}
The crystal structure of the ordered full-Heusler alloy CMS
consists of four $\emph{fcc}$ sublattices with Co atoms at (1/4,
1/4, 1/4) and (3/4, 3/4, 3/4), the Mn atom at (0, 0, 0), and the
Si atom at (1/2, 1/2, 1/2 ) in Wyckoff coordinates. These alloys
are prone to antisite disorder. A complete disorder involving the
mixing of all site, results in the $\textit{A}$2 structure type
with reduced symmetry $Im\bar3m$. The most frequently occuring
disorder type is $\textit{B}$2, where the Mn and Si atoms at
4$\textit{a}$ and 4$\textit{b}$ positions are interchanged
randomly resulting in $Pm\bar 3m$ symmetry. This type of disorder,
in which the Co positions are not affected, changes the degree of
spin polarization marginally.\cite{Kudryavtsev} Additionally,
DO$_3$ disorder can take place, which consists of random exchange
between Co and Mn atoms. Figure 1(a) shows the $\theta$-2$\theta$
XRD profiles of CMS films annealed at different temperatures. The
as-deposited film does not show any resolvable Bragg peaks
suggesting a nano-crystalline nature. However, with the increase
in $T_A$, the diffraction lines appear and their intensity and
sharpness become prominent. The diffraction profile of
600$^\circ$C annealed film shows intense $(002)$ and $(004)$
reflections of the cubic phase. The ratio of the intensities of
the $(002)$ and $(004)$ Bragg peaks, which is a measure of the
degree of crystallographic order on the Co sites,\cite{webster}
also increases with $T_A$. Figure 1(b) shows the GIXRD spectra for
600$^\circ$C annealed film taken to check the presence of $(111)$
superlattice, that governs the ordering of the Mn and Si
sublattices and $(022)$ fundamental diffraction line, which
confirms the presence of $\textit{L}2_1$ ordering in the films.
This is also earlier confirmed by taking the $\varphi$ scans for
$(111)$ and $(022)$ peaks of the CMS film.\cite{Pandey} The
$\varphi$ values for the $(111)$ and $(022)$ planes of the CMS
film are shifted by 45$^0$ with respect to those of the MgO
planes, clearly demonstrates the fourfold symmetry of cubic phase
of 600$^\circ$C annealed film. Rocking curves of the films have a
full width at half maximum (FWHM) of 0.4$^0 - 1.4^0$, indicating a
high-quality epitaxial growth. A typical rocking curve about
$(004)$ CMS peak is shown in Fig. 1(c). The FWHM of these peaks
decreases with the increase in $T_A$, which suggests that the
(00$\ell$) texture improves with increasing $T_A$. Figure 1(d)
shows the variation of $B2$ and $\textit{L}2_1$ ordering parameter
with $T_A$ calculated using the formulation given by Takamura
\textit{et al}.\cite{Takamura} These parameters give the measure
of the atomic ordering in the ordered/diordered state of the
Heusler alloys films.
\begin{figure}[h]
\begin{center}
%\vskip -1.5cm
%\abovecaptionskip -1cm
\includegraphics [width=8cm]{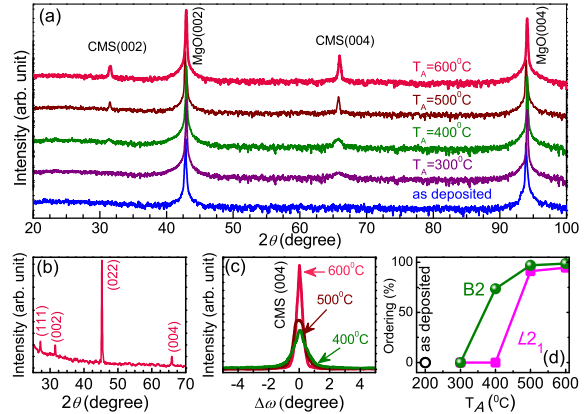}%
\end{center}
\caption{\label{FIG. 1.} (a) The $\theta$-2$\theta$  X-ray
diffraction patterns of 40 nm thick CMS films grown on MgO and
subsequently annealed at different temperatures. The ($002)$ and
$(004)$ peaks of CMS along with those of the substrates are
marked. (b) GIXRD curve of CMS/MgO film annealed at 600$^\circ$C
confirming the presence of $\textit{L}2_1$ ordering in the film.
(c) The rocking curves for $(004)$ peaks of CMS films for
different $T_A$. (d) The variation of $B2$ and $\textit{L}2_1$
structural ordering parameter with $T_A$.}
\end{figure}

The SEM micrographs of CMS films processed under different
conditions are presented in Fig 2(a-c). These surfaces are devoid
of any particulate matter, which is commonly seen in PLD grown
films if the conditions are not optimized. We also note that the
films become smoother as $T_A$ increases. The non-contact AFM
images of the above mentioned films are shown in Fig. 2(d-f). The
roughness obtained from AFM scans and XRR measurements are
compared in Fig. 2(g) (upper panel), while the lower panel shows
the variation of grain size with $T_A$ estimated by using
Debye-Scherrer method. As the $T_A$ increases the grain size
increase to attain $(00\ell)$ orientation.

\begin{figure}[t]
\begin{center}
\vskip 0.5cm
%\abovecaptionskip -1cm
\includegraphics [width=8cm]{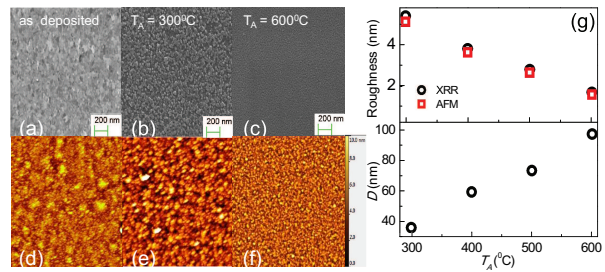}%
\end{center}
\caption{\label{FIG. 2.} (a-c) The SEM micrographs of CMS/MgO
films for as-deposited, 300$^\circ$C and 600$^\circ$C annealed
film. (d-f) The non-contact mode AFM images of the same films as
shown in (a-c). The scan area is 5$\times$5 $\mu$m$^2$. (g) The
variation of film roughness (upper panel) and grain size (lower
panel) with $T_A$.}
\end{figure}

In summary, these PLD grown films show three characteristically
different structural states; (i) the as-prepared films having
nanocrystalline phase with strong site disorder, (ii) the
nanocrystalline phase with a substantial degree of site order
inside the grains, and (iii) the crystalline phase with high
degree of site order and structural long-range order.

\subsection{Evolution of magnetic order on annealing}
The magnetization ($M$) vs field ($H$) loops of the 600$^\circ$C
annealed film taken at room temperature in out-of-plane and
in-plane field directions are shown in Fig. 3(a). We clearly see
that the films are magnetically soft with in-plane magnetization
easy axis, coercive field ($H_C$) $\approx$ 7 mT, and almost
square loops reaching saturation within 300 mT. The saturation
magnetic moment extracted from these data is $\approx$
4.78$\pm$0.19 $\mu_B$/f.u., which is in accordance with the
theoretical predictions.\cite{Galanakis} The $H_C$ of the films
depends on the internal stresses, grain size, crystal structure,
magnetic inhomogeneities, surface roughness, and thickness. The
as-deposited film shows a much higher $H_C$ compared to annealed
films. The variation of $H_C$ with $T_A$ is shown in Fig. 3(b).
The $M_S$ of the films increases with the increasing $T_A$ and
approaches a maximum at 600$^\circ$C for $\textit{L}2_1$ ordered
state [see Fig. 3(c)]. Clearly, increase in the structural
disorder
($\textit{L}2_1$$\rightarrow$$B2$$\rightarrow$$A2$$\rightarrow$
amorphous state) causes reduction in $M_S$ values. By using the
$M_S$ determined from hysteresis loops and out-of-plane saturation
field $H_K$, one can calculate the perpendicular magnetic
anisotropy energy density,
\begin{eqnarray}
K=\int{M\cdot dH = \int{\left({M^\bot- M^{||}}\right)\cdot dH}}
\end{eqnarray}

which can be further simplified as $K= M_SH_K/2$. We obtained a
$K$ value of -5.82$\times 10^4$ J/m$^3$ for 600$^0$C annealed
film, which is close to that obtained from ferromagnetic resonance
measurement on such films.\cite{Pandey} The variation of $K$ as a
function of $T_A$ is plotted in Fig. 3(d). We observe the $K$ to
be negative for all the films, implying an in-plane easy axis,
which is consistent with the data shown in Fig. 3(a).

\begin{figure}[h]
\begin{center}
%\vskip -1.5cm
%\abovecaptionskip -1cm
\includegraphics [width=8cm]{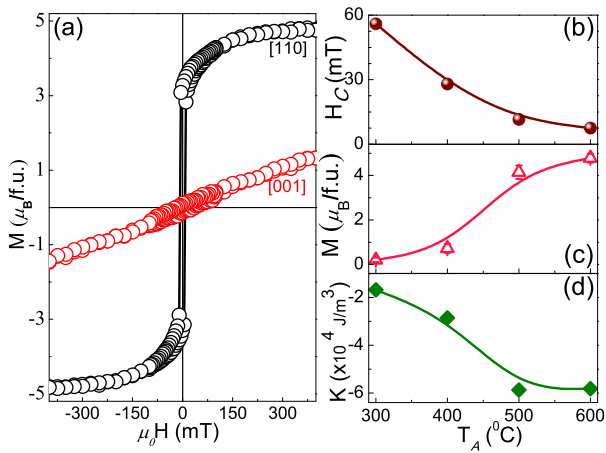}%
\end{center}
\caption{\label{FIG. 3.} (a) The room temperature magnetic
hysteresis loops with external magnetic field $\mu_0H$ applied
along [110], and [001] CMS films annealed at 600$^\circ$C. The
variation of (b) coercivity, (c) saturation magnetization, and (d)
perpendicular magnetic anisotropy constant with annealing
temperature.}
\end{figure}

\subsection{Reflection of atomic ordering on electronic transport}
\subsubsection{Longitudinal resistivity $(\rho_{xx})$}

Now we describe the electrical transport in these PLD grown CMS
films. Figure 4(a) shows the temperature dependence of
longitudinal ($\rho_{xx}$) in zero field for the as-deposited
films as well as for those annealed at 300, 400, 500, and
600$^\circ$C. In the as-deposited state, the temperature
coefficient ($d\rho/dT$) of resistivity is negative in the
temperature range 5 K to 300 K with a residual resistivity ratio
[$r = \rho_{300K}/\rho_{5K}$] of 0.8. This behavior is very unlike
of a normal metal. The metallic nature of the films improves on
annealing and a value of $r$ $\approx$ 1.5 is achieved after the
heat treatment at 600$^\circ$C, which is comparable with earlier
reported values of $r$ for CMS films.\cite{Raphael} The
enhancement of $r$ with the $T_A$ is a result of the improvement
in crystallographic ordering and, hence, a reduction of defect
concentration in the film. The dependence of $r$ and $\rho_{xx}$
at 5 K with $T_A$ are shown in the inset of Fig. 4(a). These data
reveal that a distinctly metallic transport sets in only after
annealing at $T_A$$>$ 300$^\circ$C. In the case of disordered
metals the resistivity becomes temperature independent when the
elastic mean free path approaches the interatomic distances. This
sets a limit on $\rho_{xx}$ $\approx$ 150$\mu\Omega$-cm, the
so-called Mooji criterion,\cite{Mooij} often satisfied by metallic
glasses, although large deviations are seen which suggests
non-universality of this criterion. The resistivity of these films
derives contributions from (i) the massive atomic disorder almost
to the limit of being amorphous in as-grown films, (ii)
transparency of grain boundaries whose number density decreases
with increasing $T_A$, and (iii) the presence of antisite
disorder. We believe that the huge changes in resistivity after
annealing are primarily due to crystalline nature of the films and
concomitant disorder at the large number of grain boundaries. We
have examined our data using the grain boundary model proposed by
Reiss \textit{et al.},\cite{Reiss} which can be viewed as a
quantum correction to the classical Drude conductivity. This model
attributes the reduced conductivity of the granular media to the
number density and quantum transparency of grain boundaries
crossed by an electron during two successive scattering events. In
such scenario the effective mean free path is expressed as:
\begin{figure}[h]
\begin{center}
%\vskip -1.5cm
%\abovecaptionskip -1cm
\includegraphics [width=8cm]{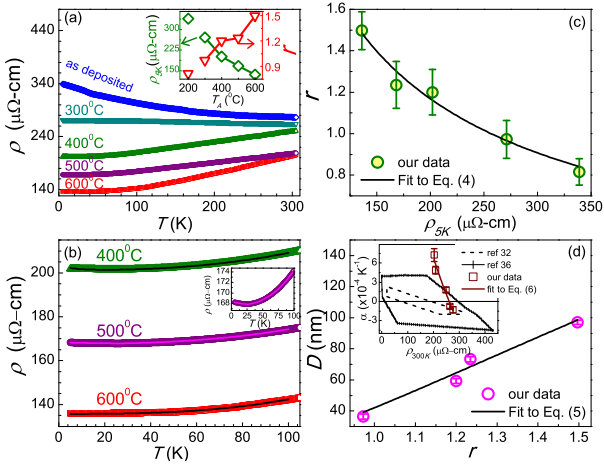}%
\end{center}
\caption{\label{FIG. 4.} (a) The resistivity of as-deposited film
as well as the films annealed at various temperatures.  The inset
shows the variation of $\rho_{5K}$ and $r$ with $T_A$. (b) The
$\rho(T)$ for the films annealed at 400, 500, and 600$^\circ$C
with their fits upto 100 K according to Eq. (7) shown by solid
lines. The inset shows the low temperature upturn for 500$^\circ$C
annealed films. (c) The $r$ as a function of $\rho_{5K}$. The
continuous line represents the fit to experimental data using Eq.
(4). (d) The $D$ obtained from AFM measurements as a function of
$r$ with their fits using Eq. (5). The inset shows the variation
of $\alpha_{300K}$ with $\rho_{300K}$ with the fit according to
Eq. (6).}
\end{figure}

\begin{eqnarray}
L_{eff}  = L\gamma^{L/D}
\end{eqnarray}

where $L$ is the mean free path in the absence of granularity,
$\gamma$ is the transmission probability of the electrons through
the grain boundary, and $D$ is the grain size. The resistivity is
thus written as:\cite{Nigro}
\begin{eqnarray}
\rho (T) = \rho _\infty  (T)\exp \left[ {\frac{{A\kappa}}{{D\rho
_\infty  (T)}}} \right]
\end{eqnarray}

where $A$ = -ln$\gamma$, $\kappa$ = $mv_F/ne^2$,
$\rho_{\infty}(T)$ = $\kappa/L$. The quantities $m$, $v_F$, and
$n$ are the mass, Fermi velocity, and density of conduction
electrons, respectively. From Eq. (3), we can easily write the
following relation:

\begin{eqnarray}
r = r_\infty  \left[ {\frac{{\rho (5K)}}{{\rho _\infty  (5K)}}}
\right]^{(\frac{1}{{r_\infty  }} - 1)}
\end{eqnarray}
where $r_{\infty} = \rho_{\infty}(300K)/\rho_{\infty}(5K)$. In
Fig. 4(c), the values of $r$ for different films are reported as a
function of $\rho_{5K}$ along with the fitting of the data by
using Eq. (4) with $r_{\infty} =(2.6\pm 0.5)$ and
$\rho_{\infty}(5K) =(53.6\pm9.3) \mu\Omega$-cm are well compared
to the values obtained for single crystal CMS.\cite{Lance} A high
value of intra-grain room temperature resistivity $\rho _\infty
(300K)=140 \mu\Omega$-cm compares in any case fairly well with the
values reported in the literature.\cite{Kartik} We can also find a
relation between $r$ and $D$ from Eq. (3) as follows:

\begin{eqnarray}
\frac{r}{{r_\infty  }} = \exp \left[ {\frac{{A\kappa}}{{D\rho
_\infty (5K)}}} \right]^{(\frac{1}{{r_\infty  }} - 1)}
\end{eqnarray}

Figure 4(d) shows the fit of the experimental data by the above
expression, which gives the value of $A\kappa$
=(7.06$\pm$0.8)$\times$ 10$^{-5}$ $\mu\Omega$-cm$^2$. Since we
have $\kappa$= (3.55$\pm$0.62)$\times$10$^{-4}$ $\mu\Omega$-cm$^2$
for our CMS films, we get $\gamma= 0.82$ and hence a fairly strong
coupling between the grains comes out even though the films have
negative thermal coefficient of resistivity (TCR). The TCR is
defined as $\alpha(T)=(1/\rho(T))/(d\rho(T)/dT)$, which can be
written as:\cite{Nigro}
\begin{eqnarray}
\alpha (T) = \alpha _\infty  (T)\left[ {1 - \ln \left( {\frac{{\rho (T)}}{{\rho _\infty  (T)}}} \right)} \right]
\end{eqnarray}
where $\alpha _\infty(T) = \frac{1}{{\rho _\infty  (T)}} {\frac{{d\rho _\infty  (T)}}{{dT}}}$.

The $\alpha_{300K}$ calculated by fitting the data with Eq. (6),
gives $\rho_{300K}$= (96.1$\pm$4.8)$\mu\Omega$-cm and
$\alpha_{\infty}$(300K)= (25.7$\pm$3.2)$\times10^{-4} K^{-1}$. Our
values of $\alpha_{300K}$ for different films agree fairly well
with the Mooij\cite{Mooij} and Tsuei\cite{Tsuei} criterion.

Now, we move to discuss the resistivity of films which have
positive TCR ($r >1$). The $\rho(T)$ of a ferromagnetic metal film
derives contributions from the defects, which is temperature
independent, as well as from electron-electron scattering (e-e),
one magnon scattering (1MS) and weak localization (WL) effects in
the disordered phase. The $\rho(T)$ below $T<100 K$ can thus be
expressed as
\begin{eqnarray}
\rho(T) = \rho _0  + A_2 T^2 + A_3 T^3 + BT^{1/2}
\end{eqnarray}
where the coefficients $A_2$, $A_3$ and $B$ corresponds to e-e,
1MS and WL processes. Here, we have ignored the $T$-linear
electron-phonon scattering term, which is expected to dominate
only for $T > 100K$. Generally, for half-metallic systems, the 1MS
process is not possible due to absence of spin down states at
$\varepsilon_F$. However, at finite temperature, the spin
fluctuations in the minority bands can make a non-zero
contribution to 1MS scattering. Furukawa has proposed that an
unconventional 1MS can be possible, which leads to $T^3$
dependence of the resistivity.\cite{Furukawa} The $BT^{1/2}$ term
in Eq. (7) emulates a low temperature upturn seen in $\rho(T)$ of
400 and 500$^\circ$C annealed film. The possible origin of this
upturn is the weak localization due to e-e interaction effects,
probably coming from the presence of impurities/disorder, or their
combined effect with lattice dynamics. We found this upturn near
25 K for 500$^\circ$C annealed film as shown in the inset of Fig.
4(b) and this minimum is not affected by the application of
external magnetic field up to 0.3 T which negates the possibility
of the well known Kondo-effect. A similar upturn in resistivity
has been reported for Heusler alloys such as
Ni$_2$MnGe,\cite{Lund} CMS\cite{Singh1,Geiersbacha} and
Co$_2$Mn$_x$Ti$_{1-x}$Al.\cite{Aftab} To understand the effect of
different scattering processes, we have fitted the resistivity
curves below 100 K by using Eq. (7) and the fitting parameters are
summarized in Table I with the error in second decimal place. From
these parameters, one can find the relative weights of the
respective resistivity components. The dominance of e-e scattering
process is likely due to grain boundaries and dominates up to 35
K. At very low temperature, the magnetic scattering of
quasiparticles is very small or nearly zero. Hence only e-e
scattering would contribute to the resistivity the most. But, at
some higher temperature the spin fluctuations near $\varepsilon_F$
modify the electronic band structure and this may lead to 1MS
scattering. The different behaviour of $\rho(T)$ curves and large
variation in residual and room temperature resistivities and $r$
in thin films as compared to those for bulk samples\cite{Lance} of
CMS could also be due to the significant contribution from the
interface between the film and the substrate. This shows the
importance of investigating the interface effects on the
resistivity of these materials for fundamental and industrial
point of view.

\begin{table}[h]
\begin{center}\caption{Fitting results of Eq. (7).} \label{TABLE I.}
%\vskip -1.5cm
%\abovecaptionskip -10cm
\begin{tabular} {c c c c c c}
\hline \hline
 $T_A$ & $\rho_0$ & $A_2$ & $A_3$ & $B$ & $\chi^2$\\
 $^\circ$C & $\mu \Omega cm$ & $\mu \Omega cm K^{-2}$  & $\mu \Omega cm K^{-3}$  & $\mu \Omega cm K^{-1/2}$ &   \\
\hline
400&202.9  & 9.31$\times$10$^{-4}$ & 9.98$\times$10$^{-7}$ & -0.421 & 0.9999\\

500&169.1  & 7.51$\times$10$^{-4}$ & 9.97$\times$10$^{-7}$ & -0.333 & 0.9999\\

600&135.8  & 6.34$\times$10$^{-5}$ & 6.01$\times$10$^{-6}$ & 0      & 0.9999\\
\hline \hline
\end{tabular}
\end{center}
\end{table}

\subsubsection{Anomalous Hall Effect}
The Hall resistivity ($\rho_{xy}$) in ferromagnetic films provides
valuable information about dominant carrier scattering mechanism,
spin-orbit scattering, state of magnetization as well as
non-trivial spin texture and their stability in \textit{H-T} phase
space. The $\rho_{xy}$ of an inhomogeneous ferromagnet can be
expressed as:
\begin{eqnarray}
\rho_{xy} = R_0H + R_SM_Z +\rho^{THE}
\end{eqnarray}
here the first term is the contribution from the Lorentz force
experienced by the carriers in the presence of externally imposed
magnetic field $H$. In the simple free electron model, $R_0 =
1/ne$, where $n$ is the carrier concentration. The values for
coefficient $R_0$ are obtained from the slope of $\rho_{xy}$ data
at high field range. The second term is the well known anomalous
Hall effect (AHE), which derives contribution from the effect of
spontaneous magnetization on scattering and a non-zero band
average anomalous velocity transverse to the applied
field.\cite{Nagaosa} The values of $R_S$ for each temperature are
obtained by extrapolating the $\rho_{xy}$ data from the saturation
value to zero magnetic field. The phenomenon of AHE has already
been established by considering its extrinsic origin based on the
skew scattering\cite{Luttinger,Smit} and side jump
mechanisms\cite{Berger} on account of asymmetric scattering of
spin-polarized charge carriers in the presence of spin-orbit
interaction as well as the intrinsic origin, ascribed to Berry
phase of Bloch
electrons.\cite{Karplus,Luttinger,Smit,Berger,Berry} The anomalous
Hall resistivity ($\rho^{AHE}$ = $R_SM_Z$, $M_Z$ is the magnetic
moment measured along \textit{z} direction) scales with the
longitudinal resistivity ($\rho_{xx}$) as
$\rho^{AHE}\propto\rho_{xx}^{\beta}$, where the exponent $\beta$
varies from 1 to 2 depending on the strength of disorder. The
topological Hall resisttivity ($\rho^{THE}$) evolves with the
field, which affects the inhomogeneous spin texture in a
non-trivial manner; but reduces to zero at high fields beyond
saturation of magnetization.

\begin{figure}[h]
\begin{center}
%\vskip -1.5cm
%\abovecaptionskip -1cm
\includegraphics [width=8cm]{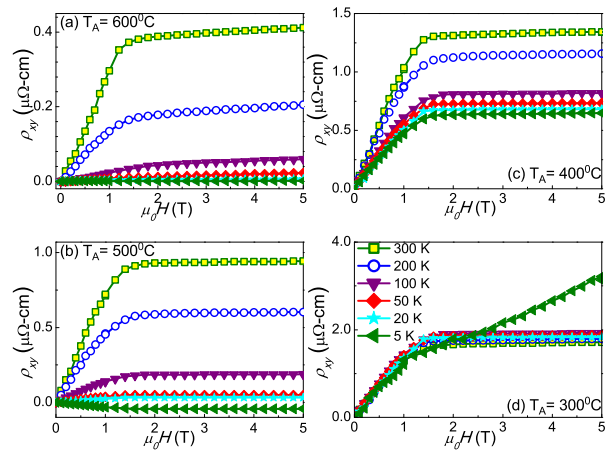}%
\end{center}
\caption{\label{FIG. 5.}Hall measurements of CMS films annealed at
various temperatures (a) 600, (b) 500, (c) 400, and (d)
300$^\circ$C.}
\end{figure}

Figure 5 shows $\rho_{xy}(H)$ data measured at different
temperatures for the CMS films. Below saturation, $\rho_{xy}$ is
governed by the AHE and increases with the increase in the sample
magnetization. Like $\rho_{xx}$, the $\rho_{xy}$ decreases with
decreasing temperature and becomes almost constant below 20 K. The
low temperature values of $\rho_{xy}$ changes with the $T_A$. For
the films annealed at 600$^\circ$C, the $\rho_{xy}$ is close to
zero, whereas, in 500$^\circ$C annealed film, the $\rho_{xy}$
gains a negative values due to DO$_3$ disorder and finally, it
attains a positive values for 400 and 300$^\circ$C annealed
samples on account of higher degree of $B$2 disorder.

Figure 6(a) shows the carrier concentration ($n$) of CMS films as
a function of temperature for different $T_A$ values. The Hall
mobility ($\mu$) = $\sigma_{xx}/ne$ remains nearly constant as
shown in Fig. 6(b). The film annealed at 600$^\circ$C is
characterized by $n$ and $\mu$ of 1.19$\times$ 10$^{29}$ m$^{-3}$
and 0.4 cm$^2V^{-1}s^{-1}$, respectively measured at room
temperature. Our values of $n$ and $\mu$ are comparable to
previously reported studies on MgO.\cite{Schneider2}
\begin{figure}[h]
\begin{center}
%\vskip -1.5cm
%\abovecaptionskip -1cm
\includegraphics [width=8cm]{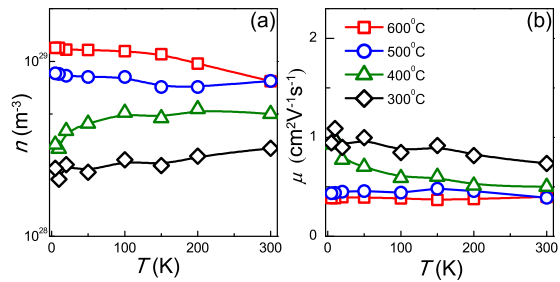}%
\end{center}
\caption{\label{FIG. 6.} The variation of (a) carrier
concentration and (b) electron mobility with temperature for CMS
films annealed at different temperatures.}
\end{figure}

The scaling between $\rho^{AHE}$ and $\rho_{xx}$ is an important
issue for better understanding of AHE. This correlation suggests
the presence of dominant scattering mechanisms in both
resistivities $\rho_{xy}$ responsible for the appearance of AHE
phenomenon caused by defects and antisite disorder.\cite{Vidal} In
general, we have,
\begin{eqnarray}
\rho^{AHE} = a\rho_{xx} + b\rho_{xx}^{2}
\end{eqnarray}
where the coefficients $a$ and $b$ give the proportions of skew
scattering, side-jump and the intrinsic mechanisms. Figure 7(a)
shows the variation of $\rho^{AHE}$ with $\rho_{xx}$ for various
films, whereas, the temperature dependence of $\rho^{AHE}$ is
shown in Fig. 7(b). This also follows as the behaviour with
respect to temperature as for $\rho_{xx}$. So $\rho^{AHE}(T)$ can
be separated into temperature-independent [$\rho^{AHE}(0)$] and
temperature-dependent [$\rho^{AHE}(T)$] resistivity below 20 K.
Hence, Eq. (9) can be rewrittten as:
\begin{eqnarray}
\rho^{AHE}(T) =\rho^{AHE}(0)+[a+2b\rho_{xx}(0)]\rho_{xx}(T)+b\rho
_{xx}^2 (T)\
\end{eqnarray}
where $\rho^{AHE}(0) = a\rho_{xx}(0)+ b\rho_{xx}^2(0)$ is the
residual anomalous Hall resistivity. By substracting $\rho^{AHE}$
values at two different temperatures, we can separate the
temperature dependent part. Hence, Eq. (10) will be reformed as:
\begin{eqnarray}
\Delta \rho^{AHE}=[a + 2b\rho_{xx}(0)]\Delta\rho_{xx}(T) +
b\Delta[\rho_{xx}^2(T)]
\end{eqnarray}
Here we have taken the difference in the resistivities with
respect to the values at 20 K and plotted in Fig. 7(c) with
fitting in according to Eq. (11). We find a linear dependence for
all the films with different slopes except for 300$^\circ$C
annealed films. The slopes of the lines decrease with increase in
$T_A$ as the skew scattering term ($a$) decreases where as the
scattering independent term ($b$) increases as shown in Fig. 7(d).
These two fitting parameters decide the dominance of the
scattering present in the films. The values for $a$ and $b$ are
(6$\pm$1)$\times10^{-3}$ and (2.39$\pm$0.26)$\times10^{-6}$
$\mu\Omega$$^{-1}$cm$^{-1}$, respectively for $L$2$_1$ ordered
600$^\circ$C annealed film and we found $[a + 2b\rho _{xx} (0)]>
b\rho _{xx} (0)$ with $\rho_{xx} (0)$ $\approx$ 135.8
$\mu\Omega$-cm. This clearly suggests the dominance of linear
scaling due to skew-scattering mechanism. The values of these
fitting parameters change very drastically as we change $T_A$. So
the structural ordering as well as magnetization in addition to
other factors like impurities in the films and their thermal
history are the important constituents which affect the scattering
mechanism in the films.
\begin{figure}[h]
\begin{center}
%\vskip -1.5cm
%\abovecaptionskip -1cm
\includegraphics [width=8cm]{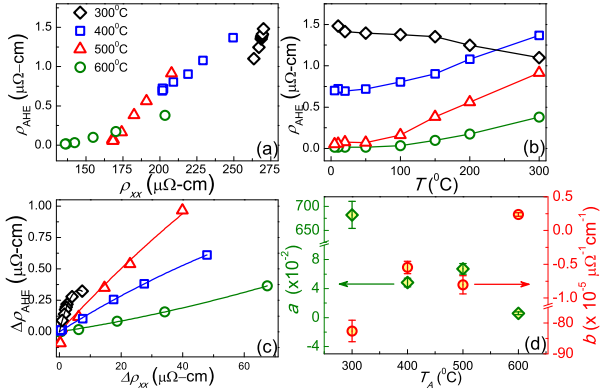}%
\end{center}
\caption{\label{FIG. 7.}(a) The variation of $\rho^{AHE}$ with
$\rho_{xx}$ for CMS films annealed at different temperatures. (b)
The temperature dependence of $\rho^{AHE}$. (c) $\Delta
\rho^{AHE}$ against $\Delta \rho _{xx} (T)$ with the fit of the
data according to Eq. (11). (d)The linear (left ordinate) and
quadratic (right ordinate) fitting parameters with varying $T_A$.}
\end{figure}

\subsubsection{Topological Hall Effect}

\begin{figure}[h]
\begin{center}
%\vskip -1.5cm
%\abovecaptionskip -1cm
\includegraphics [width=8cm]{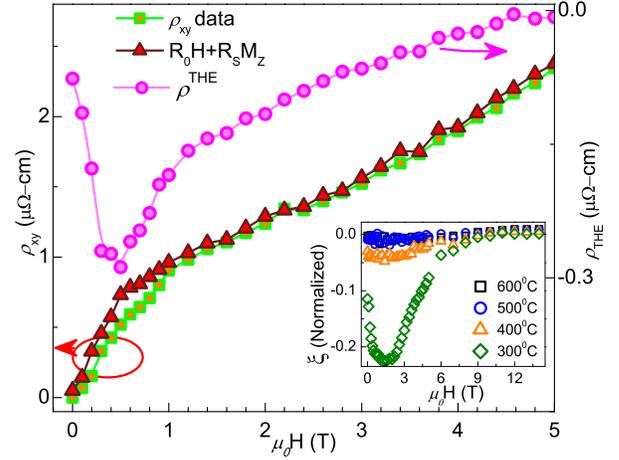}%
\end{center}
\caption{\label{FIG. 8.}Hall resistivity measured at 5 K as a
function of field for 300$^\circ$C annealed film. The inset shows
the normalized topological Hall resistivity for films annealed at
different temperature.}
\end{figure}
Now we discuss the THE, which emerges due to inhomogeneous spin
textures in the presence of magnetic field, but reduces to zero at
very high field. The film annealed at 300$^\circ$C may show a non
trivial spin character as it contains sufficient amount of
antisite disorder as well as smaller grain size, which can lead to
inhomogeneous magnetization. The canting of the magnetic spins
makes the magnetization of the 300$^\circ$C annealed films depart
from in-plane easy axis. The possibility of minority spins at
$\varepsilon_F$ also allows the scattering between spin-up and
spin-down carriers. Hence, the spin canting and presence of
minority spins at $\varepsilon_F$ may show some signatures of THE
in 300$^\circ$C annealed films at low temperature. The
$\rho^{THE}$ can be obtained by subtracting the ($R_0H+R_SM_Z$)
from $\rho_{xy}$. The experimental $\rho_{xy}$ data measured at 5
K for 300$^\circ$C annealed film along with the calculated
($R_0H+R_SM_Z$) values and $\rho^{THE}$, are shown in Fig. 8. The
$\rho_{xy}$ and ($R_0H+R_SM_Z$) increase monotonically with $H$
and coincide at high magnetic fields suggesting a diminishing
effect of $\rho^{THE}$. The contribution of $\rho^{THE}$ shows a
minima around 0.5 T. This type of behaviour has been observed in
previous experiments\cite{Kanazawa,Neubauer,Lee} and described in
theories.\cite{Yi} The inset of Fig. 8 shows a systematic
variation of normalized topological Hall resistivity
[$\xi$=$\rho^{THE}/\rho_{xy}(14T)$] for differently annealed films
to understand the effect of disorder on $\rho^{THE}$. The films
annealed at 500 and 600$^\circ$C do not show any THE due to strong
magnetic coupling, robust uniaxial in-plane anisotropy, and a
half-metallic character. As the $T_A$ reduced to 300$^\circ$C, a
minimum in the $\xi$ occurs. This observation shows that the
inhomogeneous magnetization which contribute to $\rho^{THE}$
diminishes as the films become more ordered.

\subsubsection{Anisotropic magnetoresistance}
Ferromagnetic materials show a distinct contribution $\rho_{xx}$
that depends on the direction of magnetization with respect to the
direction of the current used to measure $\rho_{xx}$. This effect,
commonly known as Anisotropic magnetoresistance (AMR), arises from
spin-orbit interaction and is of significant technological and
fundamental interest. The AMR is defined as;

\begin{eqnarray}
\frac{{\Delta \rho }}{\rho } = \frac{{\rho {}_{||} - \rho _ \bot
}}{{\rho _ \bot  }},
\end{eqnarray}

where $\rho_{||}(\rho_\bot)$ represents the resistivity when the
current flows parallel (perpendicular) to the magnetization. The
original work of Cambell \textit{et al.} attributes AMR to
$\textit{s-d}$ scattering from the conduction states and
successfully explains the AMR data of crystalline 3$\textit{d}$
transition metal alloys.\cite{Campbell} A further improvement of
the theory by Malozemoff takes into account the $\textit{s-s}$
scattering as well,\cite{Malozemoff} but these models do not
consider the spin polarization of $\textit{s}$-states. Kokado
$\textit{et al.}$ have shown that when the dominant $\textit{s-d}$
scattering process occurs between those states which have same
type of spin, the AMR ratio is negative ($\rho_{||}<\rho_\bot$),
and when the dominant scattering occurs between opposite spin
states, the sign tends to be positive.\cite{Kokado} This
prediction has a special significance for the half-metallic
systems as they have only one type of spin states at
$\varepsilon_F$, and hence, should show a predominantly negative
AMR. We have measured the AMR of our films in a field strength of
$\leq$ 0.3 T, which is sufficient to saturate the magnetization.
\begin{figure}[h]
\begin{center}
%\vskip -1.5cm
%\abovecaptionskip -1cm
\includegraphics [width=8cm]{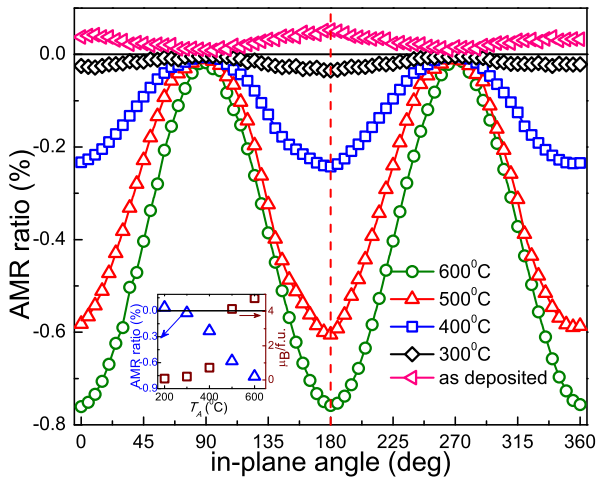}%
\end{center}
\caption{\label{FIG. 9.}Dependence of the AMR ratio on the
in-plane relative angle ($\phi$) between direction of applied
current and magnetic field for CMS films measured at 7 K. Inset
shows the variation of AMR ratio and magnetic moment with $T_A$.}
\end{figure}

Figure 9 shows the dependence of the AMR ratio on the angle $\phi$
between the current and external field for CMS films measured at 7
K, where the AMR ratio is defined as
[$(R(\phi)-R_{\bot})/R_{\bot}$], where $R_{\bot}$ corresponds to
$\phi= 90^\circ$. The obtained values of AMR ratios are comparable
with the earlier reports on mangnites \cite{Ziese,Favre} and
full-Heusler alloy Co$_2$Fe$_x$Mn$_{1-x}$Si.\cite{Yang} Our data
show a twofold symmetry for all the films with minimum occurring
at $\phi$= 0$^\circ$ and 180$^\circ$, showing negative AMR values.
The magnitude of the AMR increases with the $T_A$ (see the inset
of Fig. 9) and changes its sign for highly disordered as-deposited
film suggesting changes in the DOS at $\varepsilon_F$ and the
degree of spin polarization due to increasing structural disorder.
These observation indicates that even though the film annealed at
400$^\circ$C exhibits half-metallic character with $B2$ ordering,
there must be some overlap between $\varepsilon_F$ and the
minority band which decreases further on raising the $T_A$ due to
improvement of structural ordering.

\section{CONCLUSIONS}
In summary, we prepared laser ablated thin films of Co$_2$MnSi
with different degree of disorder realized by post-deposition
annealing in the temperature range of 300 to 600$^\circ$C. In the
as-deposited state, the film is composed of nano-crystalline
grains with $A$2 type disorder. The film behaves as a strongly
disordered metal with negative TCR due to the scattering from
grain boundaries and the site disorders within the grains.
Annealing improves structural ordering from disordered
nanocrystalline to $B$2 phase and finally highly ordered $L$2$_1$
phase is obtained at 600$^\circ$C. The system becomes metallic
with positive TCR for $T_A$ $>$ 300$^\circ$C. The saturation
magnetization builds up on annealing and reaches the bulk value as
predicted by Slater-Pauling rule at $T_A$ = 600$^\circ$C. The
behaviour of TCR, room temperature resistivity and residual
resistivity ratio are analyzed in the framework of a model based
on quantum transparency of grain boundaries. These results
establish three structural states in our films, the as-prepared
nanocrystalline phase with strong site disorder, an intermediate
phase with a small amount of site order inside the grains, and the
crystalline phase with long-range structural order. A systematic
study of anomalous Hall effect reveals the experimental evidence
of skew-scattering phenomena. We also observed an overtone of
topological Hall effect in our strongly disordered
nano-crystalline 300$^\circ$C annealed films which vanishes at
higher magnetic fields. The normalized value of topological Hall
resistivity ($\xi$) is almost zero for 500 and 600$^\circ$C
annealed films. We found a maximum value of $\xi$ $\approx$ 0.22
for 300$^\circ$C annealed film. The annealing temperature
dependence of the THE suggests the canting and/or texturing of
spins in the films annealed at 300$^\circ$C. An AMR ratio of
$\approx$ -0.76\% is obtained for 600$^\circ$C annealed film,
which decreases sharply as $T_A$ reduces, whereas, we found AMR of
$\approx$0.04 for as-deposited film. This small positive value of
AMR suggests disappearance of half-metallicity.

\section{ACKNOWLEDGEMENT}
This research has been supported by Council for Scientific and
Industrial Research (CSIR), Government of India. H.P. would like
to thank V. P. S. Awana for directing him to use the PPMS. He also
thanks the DST Unit on Nanosciences (IIT Kanpur) for the SEM and
GIXR measurements and acknowledges financial support from Indian
Institute of Technology Kanpur and CSIR. R.C.B. acknowledges J. C.
Bose Fellowship of the Department of Science and Technology, Govt.
of India.

\end{document}